\RequirePackage{lineno}
\documentclass[aps,reprint,twocolumn,a4paper,superscriptaddress]{revtex4-1}

\usepackage{graphicx}
\usepackage{dcolumn}
\usepackage{bm}
\usepackage{setspace}
\usepackage{amsmath}

\begin{document}

\singlespacing

\title{Phonon-driven ultrafast symmetry lowering in a Bi$_2$Se$_3$ crystal}

\author{A. A. Melnikov}
\email{melnikov@isan.troitsk.ru}
\affiliation {Institute for Spectroscopy RAS, Fizicheskaya 5, Troitsk, Moscow, 108840 Russia}
\affiliation {National Research University Higher School of Economics, Myasnitskaya ul. 20, Moscow 101000, Russia}
\author{Yu. G. Selivanov}
\affiliation {P. N. Lebedev Physical Institute RAS, Moscow, 119991 Russia}
\author{S. V. Chekalin}
\affiliation {Institute for Spectroscopy RAS, Fizicheskaya 5, Troitsk, Moscow, 108840 Russia}

\begin{abstract}

Selective excitation of coherent high-amplitude vibrations of atoms in a solid can induce exotic nonequilibrium states, in which the character of interactions between electronic, magnetic and lattice degrees of freedom is considerably altered and the underlying symmetries are broken. Here we use intense single-cycle terahertz pulses to drive coherently the dipole-active $E_u^1$ phonon mode of a Bi$_2$Se$_3$ crystal. As a result, several Raman-active modes are simultaneously excited in a nonlinear process, while one of them, having the $E_g^2$ symmetry, experiences dynamical splitting during the first two picoseconds after excitation. The corresponding angular scattering pattern is modified indicating coexistence of two phonon modes characteristic of a nonequilibrium state with a lower crystal symmetry. We observe also a short-lived frequency splitting of the original $E_g^2$ mode that immediately after excitation amounts to $\sim 25\%$ of the unperturbed value. This transient state relaxes with a characteristic time of $\sim$ 1 ps,  that is close to the decay time of the squared amplitude of the resonantly excited infrared-active $E_u^1$ mode. We discuss possible mechanisms of the dynamical splitting: nonlinear lattice deformation caused by the intense $E_u^1$ vibrations and excitation of anisotropic electronic distribution due to nonlinear electron-phonon interaction. Our data also contain an evidence in favor of the sum-frequency Raman mechanism of generation of the coherent $E_g^2$ phonons in Bi$_2$Se$_3$ excited by terahertz pulses. 

\end{abstract}

\maketitle

\section{Introduction}

Resonant excitation of infrared-active phonon modes by ultrashort laser pulses has recently emerged as a unique tool for the ultrafast control of microscopic processes in matter and for the investigation of exotic nonequilibrium states \cite{Kampfrath, Nicoletti, Basov, Hubener}. In particular, it was shown that dipole-active atomic vibrations can couple quadratically to the electron density of a solid \cite{Gierz, Kennes, Sentef}. This coupling considerably changes the character of electron-electron interaction provided the vibrations are driven by a strong coherent field. Among the effects that originate from this property are photoinduced superconductivity \cite{Fausti, Mitrano, Cavalleri} and light-induced electronic phase transitions \cite{Rini, Esposito}. Anharmonic interactions of the highly-excited phonon mode cause nonlinear lattice distortions, which can serve as a way of directional control of the crystal structure \cite{Forst, Subedi, Juraschek1, Mankowsky, Hoegen, Kozina, Xian}. Via spin-lattice and spin-orbital interactions laser-driven atomic motions affect magnetic degrees of freedom in solids and can induce magnetic ordering and spin-state transitions \cite{Nova, Juraschek2, Fechner}.

One of the materials that have recently attracted attention in this respect is Bi$_2$Se$_3$. It belongs to the class of bismuth and antimony chalcogenides, and, as several other compounds from this group, Bi$_2$Se$_3$ is a topological insulator. Such crystals possess unique electronic properties -- their surfaces host Dirac electronic states with linear dispersion and spin helicity \cite{Hasan, Qi}. Moreover, doped Bi$_2$Se$_3$ is now actively studied as a candidate for bulk topological superconductor and demonstrates properties characteristic of a nematic superconductor \cite{Fu, Yonezawa}. In addition, the crystal lattice of Bi$_2$Se$_3$ possesses specific anharmonism that causes good thermoelectric properties \cite{Tian}. 

Far infrared spectra of moderately doped Bi$_2$Se$_3$ crystals contain a relatively strong absorption peak that is not fully screened by the free carrier response and is due to the IR-active $E_u^1$ mode \cite{Dordevic}. Its frequency lies near 2 THz and intense coherent excitation of this mode is possible using powerful THz pulses that can be generated in this frequency range by well-established nonlinear optical techniques. Having in view the unique properties of Bi$_2$Se$_3$ mentioned above it is of interest to study the ultrafast dynamics of such high-amplitude coherent vibrational wavepackets and their interaction with electronic and structural degrees of freedom of the crystal.

Recently several remarkable phenomena have already been reported. Intense atomic $E_u^1$ oscillations were found to induce modulation of surface symmetry of a Bi$_2$Se$_3$ crystal \cite{Bowlan}. In our previous work we observed coherent nonlinear excitation of three Raman-active phonon modes of Bi$_2$Se$_3$ by a THz pulse and interpreted it as a result of anharmonic interaction with resonantly pumped $E_u^1$ mode \cite{Melnikov}. An evidence was presented also that surface-bulk electronic scattering is considerably inhibited in Bi$_2$Se$_3$ via pumping the IR-active phonon mode \cite{Yang}.

In the present work we report an observation of a short-lived splitting of the $E_g^2$ phonon mode of a Bi$_2$Se$_3$ crystal after excitation by a powerful THz pulse. This phenomenon reveals itself as appearance of two orthogonally polarized components of the original mode with blue- and red-shifted frequencies. The concomitant modification of angular dependence of the coherent phonon amplitude indicates lowering of the crystal symmetry and is accompanied by an anisotropic monotonic signal that we ascribe to nonequilibrium electrons excited via quadratic coupling to the THz-driven $E_u^1$ mode. 

The observed effect represents a specific example of an ultrafast structural transition induced by highly excited vibrations. In the framework of our interpretation it also bears an intriguing similarity to the physics of nematic superconductivity in electron-doped Bi$_2$Se$_3$. In that case the electronic order parameter (e.g. the superconducting gap) demonstrates two-fold rotational symmetry below a certain temperature and breaks the three-fold rotational symmetry of the crystal \cite{Fu, Yonezawa, Sun}. It is a matter of debate, whether this transition is accompanied by lattice distortion, and what could be the origin of this distortion \cite{Yonezawa, Kuntsevich1, Kuntsevich2, How, Frohlich}. The electronic nematicity is strongly coupled to the lattice degrees of freedom and it was proposed recently to enhance or suppress nematic order by driving the $E_u$ phonon mode in the FeSe superconductor \cite{Klein}. This approach, theoretically studied in \cite{Klein} for a tetragonal crystal, is somewhat analogous to the experimental setting used in our work for trigonal Bi$_2$Se$_3$.

\section{Experimental details}

The sample that we studied in the experiments was a single-crystal thin film of Bi$_2$Se$_3$ grown by the molecular beam epitaxy on the (111)-oriented BaF$_2$ substrate \cite{Kuntsevich3}. The film was covered by a layer of BaF$_2$ in order to protect it from ambient air. The thickness of both the Bi$_2$Se$_3$ film and the BaF$_2$ protective layer was $\sim$ 30 nm. Nearly single-cycle terahertz pulses were generated in a lithium niobate crystal in the process of optical rectification of femtosecond laser pulses with tilted amplitude fronts. The energy of pulses used for terahertz generation was 1.5 mJ at 800 nm wavelength, while the repetition rate was 1 kHz. This allowed us to obtain peak electric fields up to $\sim$ 0.5 MV/cm in a focused terahertz beam. Additional details of the setup can be found in our previous work \cite{Melnikov}. 

Using weak 800 nm probe pulses with variable optical path we detected ultrafast evolution of transmittance and optical anisotropy of the Bi$_2$Se$_3$ sample upon its excitation by a THz pulse. In order to measure changes of transmittance an additional reference beam was used. This beam was split from the probe beam before the sample and guided directly to the photodiode. The recorded experimental signal was $\Delta T/T=1-(I_{pr}/I_{ref})_{\mathrm{on}}\div(I_{pr}/I_{ref})_{\mathrm{off}}$, where $I_{pr}$ and $I_{ref}$ are the intensities of the probe and the reference pulses measured by photodiodes and on/off subscripts denote opened and closed pump THz beam. Optical anisotropy induced by the pump THz pulse was measured using only one beam. Before the sample it was polarized at 45$^{\circ}$ relative to the vertical direction of the THz electric field. Passing through the photoexcited Bi$_2$Se$_3$ crystal probe radiation experienced small rotation of polarization that was detected by splitting the probe beam into orthogonal polarization components and measuring their intensities $I_x$ and $I_y$. In this case the anisotropy signal was calculated as $1-(I_x/I_y)_{\mathrm{on}}\div(I_x/I_y)_{\mathrm{off}}$. All experiments were performed at room temperature and in nitrogen atmosphere. The latter condition allowed minimization of distortion of the THz pulse waveform.

\section{Results and discussion}

\begin{figure}
\begin{center}
\includegraphics{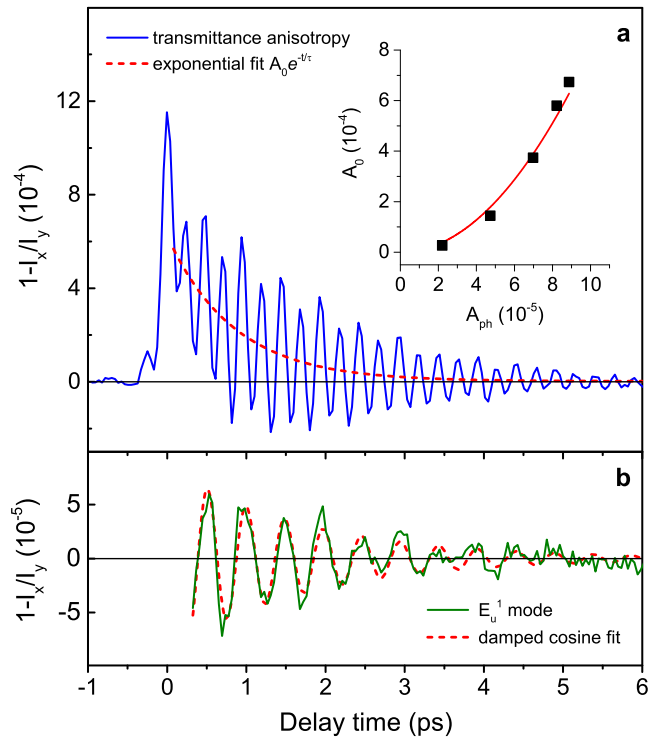}
\end{center}
\caption{\label{fig1} (a) Anisotropy of transmittance at 800 nm induced by the pump THz pulse as a function of time (solid curve). Dashed curve -- single-exponential fit of the monotonic component. The inset shows the dependence of the amplitude of the exponent on the amplitude of the $E_u^1$ oscillations. The solid line in the inset shows the parabolic fit $y=ax^2$ (b) $E_u^1$ oscillations extracted from the signal in panel (a) (solid line) and the fit by a damped cosine function (dashed line).}
\end{figure}

The transmittance anisotropy signal induced by a THz pulse with the peak electric field of $\sim$ 0.4 MV/cm is shown in Fig. 1(a). The oscillating part of the signal is caused by coherent atomic vibrations of different symmetry and was studied in detail in our previous work \cite{Melnikov}. It was shown that the spectrum of oscillations contains three peaks corresponding to $E_g^1$, $E_u^1$, and $E_g^2$ phonon modes of Bi$_2$Se$_3$. $E_g^1$ oscillations are the weakest and are indiscernible without signal analysis.  The contribution of the  $E_g^2$ mode is the largest, several times higher than that of the  $E_u^1$ mode. Since their frequencies are related approximately as 2:1 ($\sim$ 4 THz and $\sim$ 2 THz, respectively), the observed signal represents apparent harmonic oscillations with a period of $\sim$ 250 fs and a characteristic amplitude modulation.

The infrared-active $E_u^1$ mode is resonantly excited by the THz pump pulse directly, while the Raman-active $E_g^2$ mode -- via a nonlinear second-order process. In both cases the excitation is coherent, making possible the detection of atomic motion in a pump-probe experiment \cite{Dekorsy1}. Unlike Raman modes, which modulate the index of refraction of the bulk crystal, coherent $E_u^1$ vibrations affect only optical properties of the BaF$_2$/Bi$_2$Se$_3$ interface where the bulk symmetry breaks \cite{Melnikov}. The anisotropic oscillations of transmittance of the Bi$_2$Se$_3$ sample associated with the $E_u^1$ mode were extracted from the total signal in Fig. 1(a) by spectral filtering and are shown in Fig. 1(b) together with a fit by a damped cosine function. 

The monotonic part of the transmittance anisotropy signal is also rather uncommon. Monotonic (non-oscillatory) relaxation of photoinduced changes of the optical constants of semiconductors and metals usually reflects relaxation of nonequilibrium electronic distribution created by an ultrashort pump pulse. If the amplitude of a monotonic component is anisotropic relative to the polarization of the probe pulse, it indicates transient anisotropy of electronic density. If electronic properties of the studied crystal surface are isotropic in equilibrium this transient anisotropy is very short-lived with a typical decay time of less than 100 fs. Considerably longer electronic isotropization is expected in specific situations when electron-electron and electron-phonon interactions are weak, e.g. in lightly doped or intrinsic semiconductors at low excitation densities and low temperatures or in exotic cases such as nearly intrinsic graphene \cite{Konig-Otto}. 

Bi$_2$Se$_3$ crystals usually have relatively large concentration of electrons ($\sim 10^{19}$ $\mathrm{cm}^{-3}$ for our epitaxial films). If in addition the sample is at room temperature and the basal plane surface is excited and probed, one can expect very fast relaxation of photoexcited anisotropy of electronic density and of the associated optical anisotropy. Indeed, in the experiments, in which coherent phonons in Bi$_2$Se$_3$ are excited by femtosecond near infrared laser pulses \cite{Norimatsu1}, the non-oscillatory anisotropic response is represented only by a coherent artifact signal that exists only during temporal overlap of pump and probe pulses. However, in our case of THz excitation the monotonic component of the transmittance anisotropy signal is surprisingly long-lived. As illustrated by Fig. 1(a) it can be fitted with a single-exponential function of the form $A_0\exp{(-t/\tau)}$, where $\tau=0.85\pm0.15$ ps. 

A hint about the origin of this transient anisotropy can be obtained by comparing its dynamics to the dynamics of the coherent $E_u^1$ vibrations. We have extracted the corresponding waveform from the signal shown in Fig. 1(a) using spectral filtering and fitted it by a damped cosine function $A_\mathrm{ph}\exp^{-t/\tau_\mathrm{ph}}\cos(\nu t+\varphi)$ (see Fig. 1(b)). The characteristic decay time of the coherent phonon amplitude is $\tau_\mathrm{ph}=1.75\pm0.15$ ps, which is approximately two times larger than the relaxation time $\tau$ of the monotonic component of the optical anisotropy. Moreover, we performed a series of measurements varying the strength of the peak electric field of the pump THz pulse and obtained a dependence of the amplitude of the monotonic component $A_0$  on the coherent $E_u^1$ phonon amplitude $A_\mathrm{ph}$. We have found that this dependence is quadratic and can be relatively well fitted by a parabola $y=ax^2$ that is illustrated by the inset to Fig. 1(a).

\begin{figure}
\begin{center}
\includegraphics{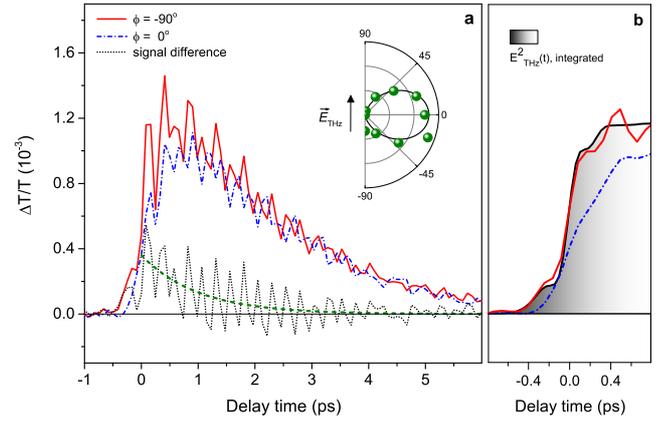}
\end{center}
\caption{\label{fig2} (a) Changes of transmittance at 800 nm induced by the pump THz pulse as a function of time for vertical ($\phi=-90^\circ$, solid curve) and horizontal ($\phi=0^\circ$, dash-dotted curve) polarizations of the probe pulse. The dotted curve is the difference between these signals and its monotonic component is indicated by the dashed line. The inset shows probe polarization dependence of the amplitude of the latter. (b) Comparison of the initial rise of the measured signals from panel (a) with the time-integrated squared field of the THz pulse (shaded area). The traces were smoothed in order to emphasize the monotonic part.}
\end{figure}

We have also studied variation of $A_0$ with the angle of probe polarization $\phi$. For that purpose we measured changes of transmittance of the Bi$_2$Se$_3$ film induced by the pump THz pulse. Decay traces obtained for vertical and horizontal polarizations of the probe pulse are shown in Fig. 2(a). As in the case of transmittance anisotropy discussed above, these $\Delta T/T$ signals consist of monotonic and oscillatory components. Temporal behavior of the former is, however, more complex than in the previous case. The main contribution to the signal is provided by the exponential relaxation with a characteristic time of $\sim$ 2.3 ps. The amplitude of this exponent is almost independent on the orientation of the probe polarization. It is thus natural to ascribe this isotropic decay to cooling of electrons heated by the pump THz pulse. The cooling occurs via interaction of hot thermalized electrons with lattice and the  characteristic time of several picoseconds is typical for this process in metals. Such interpretation and the observed relaxation rate are in agreement with previous experiments, in which ultrafast electronic dynamics in Bi$_2$Se$_3$ films was studied using near-infrared excitation \cite{Glinka}.

The hot isotropic electronic distribution is created within the duration of the THz pulse. The electric field of the pulse accelerates electrons in the vicinity of the Fermi level but the imparted momentum and energy is quickly redistributed among the ensemble via electron-electron and electron-phonon interactions. In this case the rise of the corresponding optical signal should approximately follow the function $\int_{-\infty}^{t} E_\mathrm{THz}^2(\tau) d\tau$, where $E_\mathrm{THz}$ is the electric field of the pump pulse (by analogy with Joule heating \cite{Levchuk}). As can be seen in Fig. 2(b), the initial trace of the $\Delta T/T$ signal is qualitatively similar to the time-integrated squared THz waveform if the probe pulse is vertically polarized. At the same time the increase of sample transmittance changes for horizontal probe polarization is considerably delayed. The delay is caused by an additional negative contribution of the same $\sim$ 1 ps monotonic relaxation component that is observed in the transmittance anisotropy signal (Fig. 1(a)).

Relying on these observations, in order to find the dependence of the amplitude of the faster component on the probe polarization angle $\phi$ we assume it to be zero if the probe pulse is vertically polarized (i.e. $\phi=\pm90^\circ$). Then the corresponding $\Delta T/T$ signal was subtracted from each of the traces measured at a set of angles $-90^\circ<\phi<+90^\circ$ and the obtained differential signals were fitted using the same procedure as was applied above for the monotonic part of the transmittance anisotropy signal. The obtained values are plotted in the inset to Fig. 2(a) together with a $\propto \cos^2\phi$ function, which fits the data relatively well.

Similar angular variations of ultrafast transmittance changes can be observed in semiconductors, in which interband electronic transitions are stimulated by a linearly polarized femtosecond laser pulse with a sufficiently short duration. In particular, it was shown that in GaAs \cite{Oudar} and graphene \cite{Mittendorff} this angular dependence is described by a linear function of $\cos^2\alpha$, where $\alpha$ is the angle between pump and probe polarizations (in our case $\phi=90^\circ-\alpha$). This anisotropy is caused by a specific symmetry of the matrix element of the dipole transition from valence to conduction band, due to which linearly polarized light preferentially excites electrons with wavevectors perpendicular to the direction of its polarization \cite{Mittendorff}. Via the combined action of this effect and of the Pauli blocking the resulting anisotropic electronic distribution reveals itself as anisotropy of photoinduced changes of transmittance or reflectance of the crystal and can be optically probed. Such ``optical orientation'' of electronic momenta is in fact rather general for direct band gap semiconductors and we suppose that the transient anisotropy of relative changes of transmittance of the Bi$_2$Se$_3$ crystal observed in our measurements is an indication of the short-lived anisotropic electronic distribution.

Electronic anisotropy in Bi$_2$Se$_3$ that is characterized by second angular harmonics (such as  $\propto \cos^2\phi$) is incompatible with the crystal symmetry and should induce lattice distortion. This non-uniform distortion, however, will itself cause optical anisotropy of the sample at least through modification of the electronic density of states. Thus, it could be rather difficult to separate contributions of these two effects measuring the ultrafast optical response of the crystal. Nevertheless, below we analyze the oscillating parts of  $\Delta T/T$ signals and show that these data provide a remarkable evidence of transient distortion of the Bi$_2$Se$_3$ crystal that coexists with the coherent $E_u^1$ vibrations.

Figure 3(a) shows spectra of the $\Delta T/T$ transients measured using horizontally ($\phi=0^\circ$) and vertically ($\phi=-90^\circ$) polarized probe pulses. They contain two spectral lines at $\sim$ 2.2 THz and $\sim$ 4 THz that correspond to coherent atomic vibrations of $A^1_{1g}$ and $E^2_g$ symmetry, respectively. We calculated FFT spectra for a set of $\Delta T/T$ traces measured for $-90^\circ<\phi<+90^\circ$ and obtained a spectral map shown in Fig. 3(b). It was found that the amplitude of $A^1_{1g}$ vibrations demonstrates a rather weak dependence on $\phi$, while the amplitude of $E^2_g$ oscillations is strongly modulated upon rotation of the probe polarization showing a specific angular pattern. Spectral positions and shapes of the lines also behave differently. The position of the $A^1_{1g}$ line at 2.2 THz is almost independent on $\phi$, while the $E^2_g$ line is shifted to lower frequencies at $\phi=0$ and to higher frequencies at $\phi=\pm90^\circ$. This shift is accompanied by the appearance of a wing from the low and high frequency side of the line, respectively. The $A^1_{1g}$ line also develops a wing, which, however, is much broader and appears only if the probe polarization is close to vertical (near $\phi=\pm90^\circ$). 

\begin{figure}
\begin{center}
\includegraphics{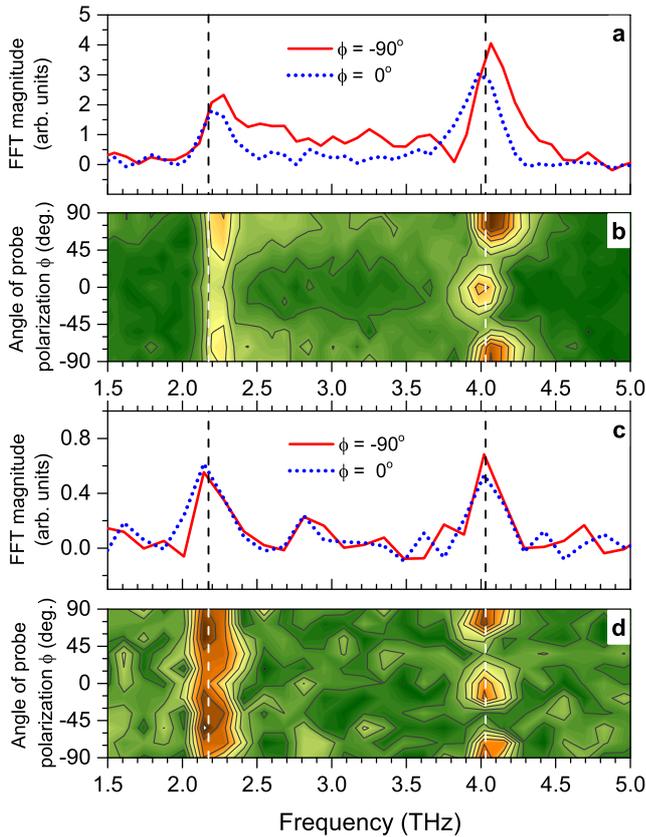}
\end{center}
\caption{\label{fig3} (a) Fast Fourier transform spectra of the $\Delta T/T$ signals measured at $\phi=0^\circ$ and $\phi=-90^\circ$. (b) Spectral map obtained by stacking the FFT spectra calculated for a set of angles of probe beam polarization from $-90^\circ$ to $+90^\circ$. (c), (d) The same type of data for truncated $\Delta T/T$ signals, containing experimental points only for $t>2$ ps.}
\end{figure}

\begin{figure}
\begin{center}
\includegraphics{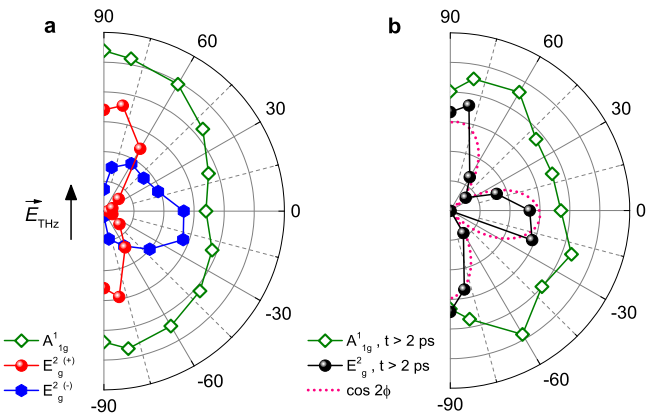}
\end{center}
\caption{\label{fig4} Dependence of the coherent phonon amplitudes on the angle $\phi$  of the probe pulse polarization for full (a) and truncated (b) $\Delta T/T$ signals. The amplitudes of $E^{2 (-)}_g$ and $E^{2 (+)}_g$ components were obtained at frequencies of 3.8 and 4.3 THz, where the bands do not overlap.}
\end{figure}

In Fig. 3(c) and Fig. 3(d) shown are the spectra of  ``truncated''  $\Delta T/T$ signals, which contain experimental points only for delay time values larger than 2 ps. These spectra are characterized by a lower signal-to-noise ratio than the original ones due to the damping of oscillations. Nevertheless, they demonstrate well that the corresponding spectral positions and profiles of both the 2.2 THz and 4 THz spectral lines almost coincide at longer delay times, when coherent atomic vibrations are unaffected by primary transient processes. In this case there is no pronounced spectral wings and the overall amplitude and angular patterns look as expected for a pump-probe measurement with excitation by a femtosecond pulse of rather low intensity. In Fig. 4(b) we plotted angular dependences of these ``unperturbed'' $A^1_{1g}$ and $E^2_g$ amplitudes extracted from the array of data shown in Fig. 3(d).

In the experiments, in which lattice dynamics is studied via Raman scattering of continuous wave laser light, the observed polarization dependence of the intensity of scattered light is associated with crystal symmetry using the corresponding Raman tensor. After certain modifications this approach can be applied to coherent lattice vibrations excited and detected by ultrashort laser pulses with a duration of less than the period of studied phonon modes. Coherent atomic vibrations reveal themselves as periodic oscillations of optical constants of a crystal through modulation of the optical susceptibility $\chi$. In the case of transmittance $\Delta T=\frac{\partial T}{\partial \chi}\frac{\partial \chi}{\partial Q}Q$, where $Q$ is the atomic displacement coordinate \cite{Dekorsy1}. The factor $\frac{\partial \chi}{\partial Q}$ is the first order Raman tensor that defines a certain dependence of the apparent amplitude of oscillations on the angle of probe polarization $\phi$,  which can be evaluated as $\vec{E}^t(\frac{\partial \chi}{\partial Q})\vec{E}^i$, where $\vec{E}^i$ and $\vec{E}^t$ are the incident and transmitted probe fields \cite{Dekorsy2}. The in-plane Raman tensors for the $A^1_{1g}$ and $E^2_g$ modes of Bi$_2$Se$_3$ can be chosen as $\begin{pmatrix} a & 0 \\ 0 & a \end{pmatrix}$ and $\begin{pmatrix} c & 0 \\ 0 & -c \end{pmatrix}$, respectively \cite{Richter}, while both probe field vectors are proportional to $\begin{pmatrix} \cos\phi \\ \sin\phi \end{pmatrix}$. Thus, coherent $A^1_{1g}$ and $E^2_g$ phonon amplitudes should behave as $\cos^2\phi+\sin^2\phi$ and $\cos2\phi$, respectively. Indeed, taking into account lower signal-to-noise ratio for truncated signals, the observed variations of the $A^1_{1g}$ amplitude can be considered as approximately isotropic, while the dependence of the $E^2_g$ amplitude on $\phi$ can be relatively well fitted by a $\propto \cos2\phi$ function (see Fig. 4(b)). This is in agreement with previous studies of coherent phonons in crystals with similar Raman tensors for $A$ and $E$ phonon modes \cite{Yee, Ishioka, Norimatsu2}. 

Polarization dependence of coherent phonon oscillations in the full $\Delta T/T$ traces demonstrates a different behavior. As we noted above, in the case of the $E^2_g$ mode not only its amplitude but also spectral position and line shape are strongly modulated by the variation of $\phi$. This behavior can be naturally explained if we suppose that the Bi$_2$Se$_3$ crystal experiences a short-lived non-uniform distortion, which is most pronounced during the first 2 ps after THz excitation. This distortion lowers the symmetry of the crystal and causes splitting of the phonon modes with in-plane atomic displacements. In the case of  $E^2_g$ mode two new modes appear instead of the original one: $E^{2 (-)}_g$ and $E^{2 (+)}_g$ with lower and higher frequencies, respectively. The frequency shifts are likely caused by the fact that the distortion changes interatomic distances and thus the bond stiffness. Since crystal symmetry is lowered upon the distortion, the low- and high-frequency components of the $E^2_g$ mode cannot be classified as $E_g$ modes. However, here we denote them as $E^{2 (-)}_g$ and $E^{2 (+)}_g$ to emphasize their origin. We note that a very similar behavior of an $E_g$ mode was observed recently in a MoS$_2$ crystal with the same D$_{3d}$ symmetry as in Bi$_2$Se$_3$ under application of static stress \cite{Lee}. In contrast to that case in our experiments the distortion and the observed frequency splitting are dynamic.

In order to evaluate the dependence of the $E^{2 (-)}_g$ and $E^{2 (+)}_g$ amplitudes on $\phi$ we selected the columns of the data array from Fig. 3(b) that correspond to frequencies of 3.8 and 4.3 THz located in the low and high frequency wings of the bands. The obtained angular patterns are plotted in Fig. 4(a). It can be seen that these polarization dependences do not follow the $\propto \cos 2\phi$ function but are rather roughly proportional to $\cos^2\phi$ and $\sin^2\phi$. Taking into account the formalism used above to describe anisotropy of detected coherent phonon amplitudes we suppose that in order to generate such angular patterns the in-plane Raman tensors of the $E^{2 (\pm)}_g$ modes should have the form $\begin{pmatrix} a & 0 \\ 0 &  b \end{pmatrix}$, where $a>>b$ or $b>>a$. The crystal structure of an undistorted Bi$_2$Se$_3$ crystal is trigonal and the in-plane Raman tensor for $A_g$ modes has the form $\begin{pmatrix} a & 0 \\ 0 &  a \end{pmatrix}$. The first crystal system of a lower symmetry, for which Raman tensors of $A$ modes contain non-equal diagonal elements is orthorhombic. Thus, we can suppose that the symmetry of the observed short-lived distorted state of Bi$_2$Se$_3$ is orthorhombic or lower. 

\begin{figure}
\begin{center}
\includegraphics{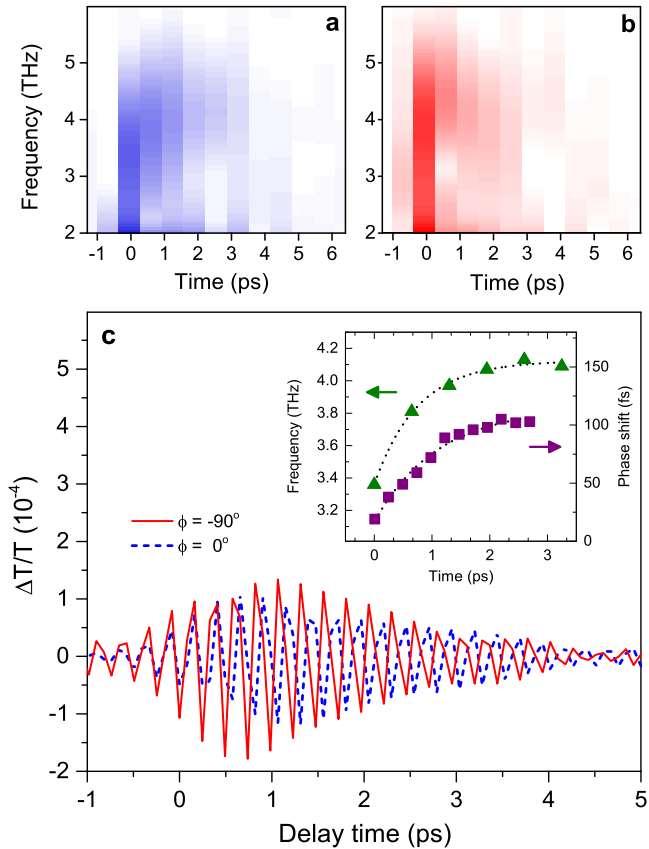}
\end{center}
\caption{\label{fig5} (a), (b) Spectral maps showing temporal evolution of the 4 THz band obtained applying short time Fourier transform to $\Delta T/T$ signals measured with horizontal and vertical probe polarization, respectively. (c) Oscillations within the 4 THz band filtered out from the same two traces. The inset illustrates temporal variation of the instantaneous frequency of $E^2_g$ oscillations in the $\Delta T/T$ signal detected with horizontally polarized probe pulse, and the phase shift between oscillations shown in panel (c).}
\end{figure}

The observed splitting of the $E^2_g$ mode in the first 2 ps after excitation by the THz pulse implies a transient character of the frequency shifts of the resulting components $E^{2 (\pm)}_g$ and a dependence of their instantaneous frequencies on time. In order to better visualize this effect we compare oscillations within the 4 THz line filtered out of the two original $\Delta T/T$ traces measured with vertically and horizontally polarized probe pulses. As follows from Fig. 5(c) the oscillations are nearly in phase at zero delay time but gradually the phase difference increases and reaches maximum by $\sim$ 2 ps when the oscillations are almost out of phase. 

Additionally we performed short time Fourier transform of the same two $\Delta T/T$ traces with the results shown in Fig. 5(a) and Fig. 5(b). This procedure implies calculation of fast Fourier transforms of sections of a measured trace contained within a window of a certain length, which is translated along the trace. The series of spectra obtained in such a way is then plotted in the form of a map that illustrates temporal evolution of the spectral lines. Here we used the Blackman window with the actual length of 1 ps and the corresponding shape-specific length of $\sim$ 0.4 ps. The overlap of adjacent windows was 0.5 ps. This combination of parameters allowed us to achieve an acceptable balance between the temporal and spectral resolution. 

The obtained spectral maps clearly demonstrate picosecond relaxation of positions of the $E^{2 (-)}_g$ and $E^{2 (+)}_g$ spectral lines towards ``quasi-equilibrium'' $E^2_g$ frequency. The initial instantaneous frequency of $E^{2 (-)}_g$ oscillations detected with horizontally polarized probe pulses reaches $\sim$ 3.4 THz, while for the $E^{2 (+)}_g$ component measured with vertical probe polarization it is roughly 4.5 THz. In the latter case (Fig. 5(b)) the spectral map is distorted due to the broad wing of the 2.2 THz line and it is difficult to evaluate the character of frequency relaxation. However, in the case of horizontally polarized probe (Fig. 5(a)) the instantaneous maximum of the spectral line demonstrates a distinct exponential-like behavior. We have found that the corresponding set of points can be relatively well fitted by a single exponential function with a characteristic time of $\tau_{\Delta\nu}=0.75\pm0.08$ ps. These data are plotted in the inset to Fig. 5(c) together with the values of time-varying phase shift of oscillating components shown in the main panel. The evolution of the phase shift is also exponential with a characteristic time of $\tau_{\Delta\varphi}=1.1\pm0.15$ ps. Thus, the characteristic relaxation time of the observed frequency splitting of the $E^2_g$ mode is the same (within experimental error) as that of the monotonic anisotropic component ($\tau$), while both these values coincide with the decay time of the squared coherent $E^1_u$ phonon amplitude ($\tau_{ph}/2$).

 In an attempt to explain this simultaneity we suppose that the observed short-lived distortion of the Bi$_2$Se$_3$ crystal is caused by the nonequilibrium anisotropic electronic density, which in turn is induced via quadratic coupling to the highly excited $E^1_u$ mode \cite{Gierz, Rini, Klein}. The equality $\tau\approx\tau_{ph}/2$ is possible if the characteristic time of electronic anisotropy relaxation is much less than $\tau$ but not too small in comparison with the half-period of $E^1_u$ vibrations (250 fs). An estimate of $\sim$ 100 fs is a realistic value for the timescale of electronic isotropization in Bi$_2$Se$_3$. In this case coherent $E^1_u$ oscillations will sustain transient electronic anisotropy and thereby the lattice distortion as long as their squared amplitude is nonzero. 

At the same time we believe that direct generation of crystal distortion via nonlinear coupling of intense coherent $E^1_u$  atomic vibrations with other lattice degrees of freedom is unlikely for the following reasons. The most efficient process is expected to be the coupling of $E^1_u$ and $E^2_g$ modes via three-phonon interaction $\propto Q^2_{E_u}Q_{E_g}$ \cite{Juraschek1}. Since THz-excited $E^1_u$ vibrations are coherent, this interaction should induce non-oscillatory directional displacement of the atomic $Q_{E_g}$ coordinate or, in other words, ``rectification'' of oscillations, using the optical analogy. This displacement should in principle reveal itself in the transient anisotropy response of Bi$_2$Se$_3$ as a picosecond monotonic component. However, as we have shown in our previous work \cite{Melnikov} by modelling the $E^1_u$ and $E^2_g$ modes as coupled classical oscillators, this directional displacement is much smaller than the amplitude of coherent $E^2_g$ oscillations. Thus, the relation between amplitudes of the corresponding monotonic and oscillating components of the anisotropy response detected in a pump-probe measurement should be the same since both are driven by the $Q_{E_g}$ coordinate. In our experiments we observe the opposite: the two amplitudes are comparable. Moreover, if the $Q_{E_g}$ displacement was so large that it caused mode splitting, then the evolution of coherent $E^2_g$ phonons would be strongly nonlinear along the trace. However, a simple damped oscillator response is observed after fast relaxation of the transient distortion. 

\begin{figure}
\begin{center}
\includegraphics{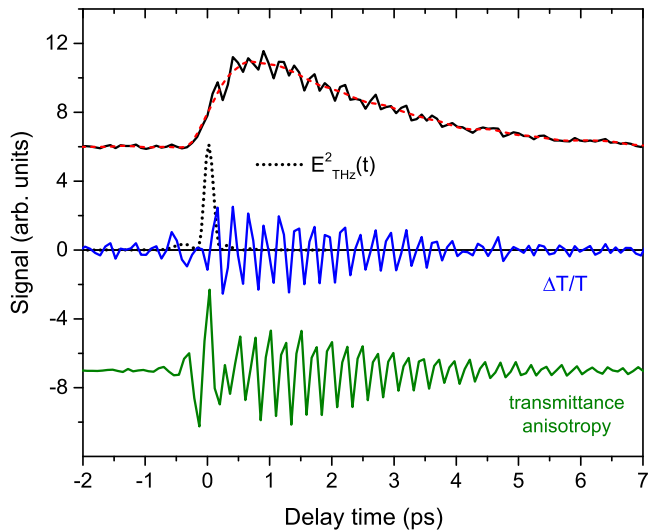}
\end{center}
\caption{\label{fig6} The upper solid curve is the $\Delta T/T$ trace measured with horizontal probe polarization. The dashed curve indicates its non-oscillating component. The middle curve is the oscillating component of the $\Delta T/T$ trace above within the 4 THz band. The squared pump THz waveform is shown with the dotted line. The bottom trace is the oscillating part of the transmittance anisotropy signal. All curves were normalized and translated appropriately for better clarity.}
\end{figure}

Additionally, the polarization-resolved data that we have obtained contain an evidence that the three-phonon interaction $\propto Q^2_{E_u}Q_{E_g}$ in Bi$_2$Se$_3$ is in fact not an effective way of displacing the $Q_{E_g}$ coordinate. The observed very specific dependence of the instantaneous frequency of the coherent $E^2_g$ vibrations on probe polarization and time has a rather unexpected consequence. As mentioned above, $E^2_g$ oscillations in traces measured with vertically ($(\Delta T/T)_y$) and horizontally ($(\Delta T/T)_x$) polarized probe pulses are initially in phase. Since THz-induced changes of sample transmittance are rather small, it can be shown using the formulae from experimental details section that the transmittance anisotropy signal is very close to $(\Delta T/T)_x-(\Delta T/T)_y$. Therefore, in phase oscillations of transmittance will effectively cancel each other, while the out of phase oscillations will be enhanced. This peculiarity explains the delayed $\sim$ 1 ps growth of the apparent amplitude of $E^2_g$ oscillations in the transmittance anisotropy response of Bi$_2$Se$_3$ to the THz pulse. This effect is emphasized in Fig. 6, where we compare oscillating components of transmittance anisotropy and of transmittance changes. Indeed, the delayed increase of the coherent amplitude of $E^2_g$ vibrations is clearly visible in the transient anisotropy signal. At the same time oscillations in the $\Delta T/T$ trace start ``immediately'' with a maximal amplitude, and this start is compatible with the duration of the squared THz waveform. We note that in our previous work \cite{Melnikov} we considered the picosecond growth of $E^2_g$ amplitude as a key evidence in favor of the lattice anharmonicity mechanism of the nonlinear generation of coherent phonons in Bi$_2$Se$_3$. Therefore our present data indicate that coherent THz excitation of at least $E^2_g$ phonon mode in Bi$_2$Se$_3$ is an impulsive process, which is probably associated with the sum-frequency Raman scattering \cite{Maehrlein}.

\section{Conclusion}

In conclusion, we have detected dynamical splitting of the $E^2_g$ phonon mode of a Bi$_2$Se$_3$ crystal pumped by a powerful single-cycle terahertz pulse. The splitting is a result of transient lattice distortion that has a characteristic relaxation time of $\sim$ 1 ps and reveals itself as appearance of red- and blue-shifted frequency components of the original $E^2_g$ mode. The initial magnitude of the frequency splitting is rather large and immediately after excitation amounts to $\sim 25\%$ of the equilibrium $E^2_g$ frequency. This ultrafast structural effect is accompanied by optical anisotropy of electronic origin, which has a two-fold rotational symmetry. In order to interpret these observations we have supposed that the anisotropic electronic density is generated through quadratic coupling with the high-amplitude coherent $E^1_u$ atomic vibrations. The non-uniform lattice distortion is then a necessary rearrangement of the crystal required to match the lower symmetry of the excited state. The observed phenomenon is the example of an ultrafast structural transition induced by intense terahertz radiation via resonant excitation of an infrared-active phonon mode. The manipulation of final phases of the frequency splitted components via variation of the pump THz pulse intensity can in principle be used as a way to control the trajectory of atomic motion of $E^2_g$ symmetry. Our polarization resolved measurements have also revealed that the nonlinear excitation of coherent $E^2_g$ phonons is in fact impulsive and can be probably described as the sum-frequency Raman scattering.

\begin{acknowledgments}

The reported study was funded by the Russian Foundation for Basic Research, project number 20-02-00989.

\end{acknowledgments}

\end{document}